\documentclass[runningheads]{llncs}
\usepackage[T1]{fontenc}
\usepackage[
   a4paper,
   pdftex,
   pdftitle={A Higher-Order Vampire},
   pdfauthor={Ahmed Bhayat, Martin Suda}
]{hyperref}
\hypersetup{final}

\usepackage{graphicx}
\usepackage{bussproofs}
\usepackage{amsfonts} 
\usepackage[binary-units=true]{siunitx}
\usepackage{todonotes}
\usepackage{textcomp}

\renewcommand{\inf}[1]{\textsc{#1}}

\newcommand{\un}[2]{U_n(#1, #2)}
\newcommand{\eq}{\approx}
\newcommand{\noteq}{\not\eq}
\newcommand{\eqOrNorEq}{\, \dot{\eq} \,}

\newcommand{\nfosubterm}[2]{#1 \langle \, #2  \, \rangle}
\newcommand{\tuple}[1]{ \overline{#1} } 
\newcommand{\typequant}{\mathsf{\Pi}}

\newenvironment{scprooftree}[1]%
  {\gdef\scalefactor{#1}\begin{center}\proofSkipAmount \leavevmode}%
  {\scalebox{\scalefactor}{\DisplayProof}\proofSkipAmount \end{center} }

\def\orcidID#1{\href{http://orcid.org/#1}{\raisebox{-1.25pt}{\includegraphics{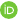}}}}

\title{A Higher-Order Vampire (Short Paper)}
\author{Ahmed Bhayat\inst{1}\orcidID{0000-0002-1343-5084} \and
Martin Suda\inst{2}\orcidID{0000-0003-0989-5800}}
\authorrunning{Bhayat and Suda}
\institute{Independent Scholar, Leicester, UK\\
\email{ahmed\_bhayat@hotmail.com} \and
Czech Technical University in Prague, Czech Republic\\
\email{martin.suda@cvut.cz}}

\begin{document}
\maketitle              
\begin{abstract}
The support for higher-order reasoning in the Vampire theorem prover has recently been completely reworked. 
This rework consists of new theoretical ideas, a new implementation, and a dedicated strategy schedule. 
The theoretical ideas are still under development, so we discuss them at a high level in this paper. 
We also describe the implementation of the calculus in the 
Vampire theorem prover, the strategy schedule construction and several empirical performance statistics.


\end{abstract}
\section{Introduction}

The Vampire prover \cite{vampire} has supported higher-order reasoning since 2019 \cite{bhayat2020combinator}.
Until recently, this support was via a translation from higher-order logic (HOL)
to polymorphic first-order logic using combinators. 
The approach had positives, specifically it avoided the need
for higher-order unification.
However, our experience suggested that for problems requiring complex
unifiers, the approach was not competitive with calculi that 
do rely on higher-order unification.
This intuition was supported by results at the CASC system competition \cite{casc}.

Due to this, we recently devised an entirely new higher-order 
superposition calculus. 
This time we based our calculus on a standard presentation of HOL.
The key idea behind our calculus is that rather than using full higher-order unification, 
we use a depth-bounded version. 
That is, when searching for higher-order unifiers, when some predefined number
of projection and imitation steps have taken place, the search is backtracked.
The crucial difference in our approach to similar approaches 
is that rather than failing 
on reaching the depth limit, we turn the set of remaining unification
pairs into negative constraint literals which are returned along
with the substitution formed until that point.
This is similar to recent developments in the field of theory reasoning \cite{bhayat2023refining}.

The new calculus has now been implemented in Vampire
along with a dedicated strategy schedule.
Together these developments propelled Vampire to first place in the THF division
of the 2023 edition of the CASC competition.\footnote{\url{https://tptp.org/CASC/29/WWWFiles/DivisionSummary1.html}}
As the completeness of the calculus is an open question
which we are working on, we have to date not published a description of the calculus.

In this paper, we describe the calculus, discuss its implementation in Vampire
and also provide some details of the strategy schedule and its formation.

\section{Preliminaries}
\label{sec:tech_background}

We assume familiarity with higher-order logic and 
higher-order unification. 
Detailed presentations of these can be found in 
recent literature \cite{bhayat2020thesis,bentkamp2023superposition,vukmirovic2021efficient}

We work with a rank-1 polymorphic, clausal, higher-order logic.
For the syntax of the logic we follow a more-or-less standard
presentation such as that of Bentkamp et al. \cite{bentkamp2023superposition}. 
Higher-order applications such as $ f \, a \, c$ 
contain subterms with no first-order equivalents such as $f$ and $f \, a$. 
We refer to these as \emph{prefix} subterms.
We represent term variables with $x, y, z$,
function symbols with $f, g, h$, and terms with $s$ and $t$.
To keep the presentation simple, we omit
typing information from our terms. 

A substitution is a mapping of variables to terms.
Unification is the process of finding a substitution $\sigma$
for terms $t_1$ and $t_2$ such that $t_1\sigma \approx t_2\sigma$
for some definition of equality ($\approx$) of interest. 
It is well known that first-order syntactic unification is decidable
and unique most general unifiers exists.
For the higher-order case, unification is not 
decidable, and the set of incomparable unifiers is potentially infinite.
A commonly used higher-order unification procedure for 
enumerating unifiers is Huet's preunification routine \cite{huet1975unification}.
Unlike full higher-order unification, preunification does not attempt
to unify terms if both have variable head symbols. 
Thus, preunification does not require infinitely branching 
rules unlike full higher-order unification \cite{vukmirovic2021efficient}.
 
The two main rules that extend first-order unification in Huet's procedure are \emph{projection} adn \emph{imitation}. 
We provide a flavour of these via an example. Consider unifying terms $s = x \, a$ and $s' = a$. In searching for a suitable instantiation of the variable $x$, we can either attempt to copy the head symbol of $s'$ leading to the substitution $x \rightarrow \lambda y. \, a$, or we can bring one of $x$'s arguments to the head position leading to the substitution $x \rightarrow \lambda y. \, y$. The first is known as imitation and the second as projetcion.
 
We use the concept of a \emph{depth$_n$ unifier}.
We do not define the term formally, but provide an 
intuitive understanding. Consider a higher-order preunification algorithm.
Any substitution formed by following a path of the unification
tree, starting from the root, that contains exactly $n$
imitation and projection steps, or
reaches a leaf using fewer than $n$ such steps,
is a \emph{depth$_n$ unifier}.
For terms $s$ and $t$, let $\un{s}{t}$ be the set of
all depth$_n$ unifiers of $s$ and $t$. 
Note that this set is finite as we are assuming preunification
and hence the tree is finitely branching.
 
For terms $s$ and $t$, for each depth$_n$ unifier $\sigma \in \un{s}{t}$,
we associate a set of negative equality literals
$C_\sigma$ formed by turning the unification pairs that remain when the
depth limit is reached into negative equalities.
In the case $\sigma$ is an \emph{actual unifier} of $s$ and $t$,
$C_\sigma$ is of course the empty set.

\begin{figure}
\centering
\includegraphics[scale=0.45]{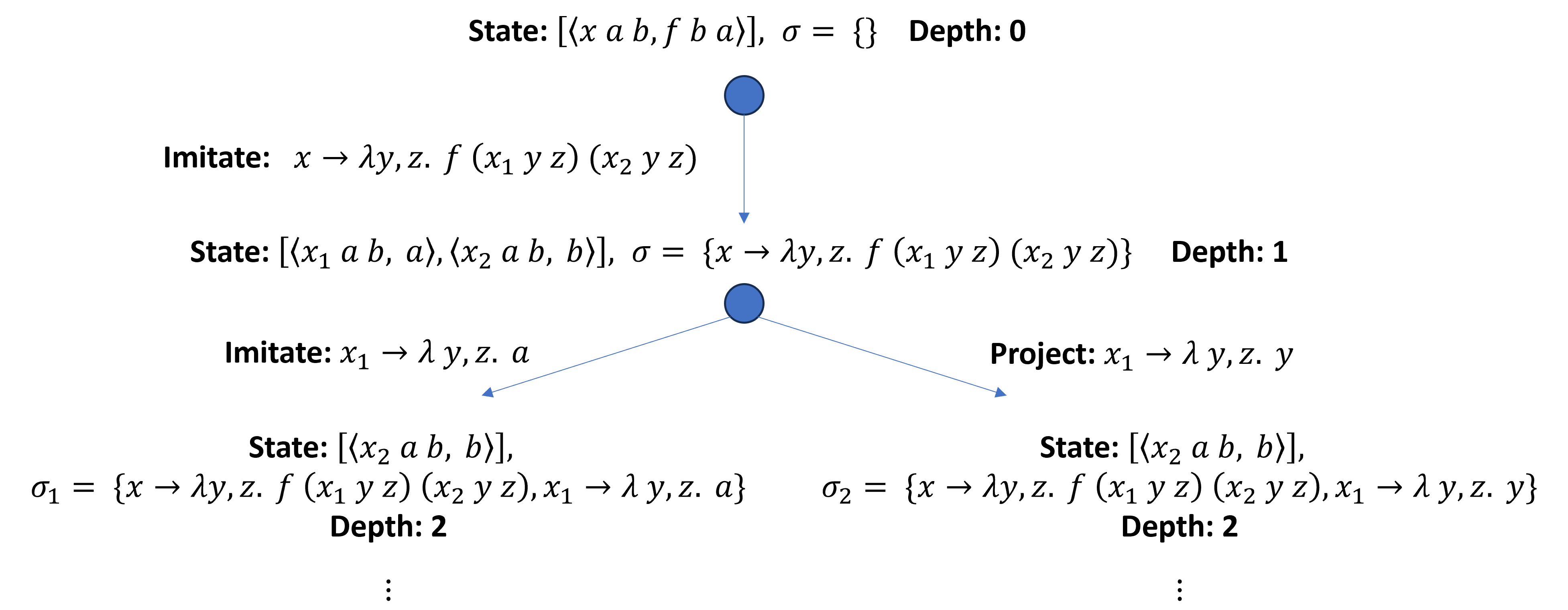}
\label{fig_unif}
\caption{Unification tree for terms $x \, a \, b$ and $f \, b \, a$}
\end{figure}

To make this clearer, consider the unification tree
presented in Figure \ref{fig_unif}.
There are two depth$_2$ unifiers labelled $\sigma_1$ and $\sigma_2$
in the figure.
Related to these, we have $C_{\sigma_1} = C_{\sigma_2} = \{ x_2 \, a \, b \, \noteq b \}$.
There are four depth$_3$ unifiers (not shown in the figure) and zero
depth$_n$ unifiers for for $n > 3$.

\section{Calculus}
\label{sec:calculus}

Our calculus is parameterised by a selection function and an ordering $\succ$. Together these give rise to the concept of literals being (strictly) $\succ$-eligible with respect to a substitution $\sigma$ \cite{bentkamp2023superposition}. 
When discussing eligibility we drop $\succ$ and $\sigma$ and rely on the context to make these clear. 
We call a literal $s \noteq t$, where both $s$ and $t$
have variable heads, a \emph{flex-flex} literal.
Such a literal is never selected in the calculus.
We present the primary inference rule, \inf{Sup}, below.

\begin{center} 

    \begin{scprooftree}{1}
        \def\defaultHypSeparation{\hskip .1in}    
        \AxiomC{$D' \vee t \eq t'$}
        \AxiomC{$C' \vee \nfosubterm{s}{u} \eqOrNorEq s'$}
        \RightLabel{\inf{Sup}}
        \BinaryInfC{$(C' \vee D' \vee \nfosubterm{s}{t'} \eqOrNorEq s' \vee C_{\sigma})\sigma$}  
    \end{scprooftree}

\end{center}

\noindent
In the rule above, we use $\eqOrNorEq$ to denote either a positive or negative equality. We use $\nfosubterm{s}{u}$  to denote that $u$ is a \emph{first-order} subterm of $s$. 
That is, a non-prefix subterm that is not below a lambda.
The side conditions of the inference are
$\sigma \in \un{t}{u}$, 
$u$ is not a variable,
$t \eq t'$ is strictly eligible in the left premise, 
$\nfosubterm{s}{u} \eqOrNorEq s'$ is eligible in the right premise, and the other standard ordering conditions.
The remaining core inference rules are \inf{EqRes} and \inf{EqFact}.

\begin{center}
  \begin{minipage}[b]{.48\textwidth}

    \begin{scprooftree}{1}
        \AxiomC{$C' \vee t \eq t' \vee s \eq s'$}
        \RightLabel{\inf{EqFact}}
        \UnaryInfC{$(C' \vee t' \noteq s' \vee s \eq s' \vee C_\sigma)\sigma$}  
        \singleLine
    \end{scprooftree}

  \end{minipage}
  \quad
  \begin{minipage}[b]{.48\textwidth}

    \begin{scprooftree}{1}
        \AxiomC{$C' \vee s \noteq t'$}
        \RightLabel{\inf{EqRes}}
        \UnaryInfC{$(C' \vee C_\sigma)\sigma $}  
        \singleLine
    \end{scprooftree}

  \end{minipage}
\end{center}

\noindent
For both rules, $\sigma \in \un{t}{s}$. 
For \inf{EqFact}, $s \eq s'$ is eligble in the premise and for 
\inf{EqRes} $s \noteq s'$ is eligble.
We also include inferences \inf{ArgCong} (see \cite{bentkamp2019lhol}), and 
\inf{FlexFlexSimp} which derives the empty clause, $\bot$, from a clause
containing only flex-flex literals.

\begin{center} 
  \begin{minipage}[b]{.48\textwidth}

    \begin{scprooftree}{1}
        \AxiomC{$C' \vee s \eq s'$}
        \RightLabel{\inf{ArgCong}}
        \UnaryInfC{$C'\sigma \vee (s\sigma) \, x \eq (s'\sigma) \, x$}         
    \end{scprooftree}
  \end{minipage}
  \quad
  \begin{minipage}[b]{.48\textwidth}
    \begin{scprooftree}{1}
        \def\defaultHypSeparation{\hskip .1in}    
        \AxiomC{$x_1 \, \tuple{s}_n \noteq x_2 \, \tuple{r}_m \vee \cdots$}
        \RightLabel{\inf{FlexFlexSimp}}
        \UnaryInfC{$\bot$}  
    \end{scprooftree}
  \end{minipage}
\end{center}

\noindent
For \inf{ArgCong}, $s \eq s'$ is eligible in the premise, 
$\sigma$ is the type unifier of $s$ and $s'$ and $x$ is a fresh variable.
In our implementation, the depth parameter $n$ is set via a user option.
In the case it is set to 0, the following pair of inferences
are added to the calculus.

     \begin{scprooftree}{1}
        \AxiomC{$C' \vee x \, \tuple{s}_n \noteq f \, \tuple{t}_m $}
        \RightLabel{\inf{Imitate}}
        \UnaryInfC{$(C' \vee x \, \tuple{s}_n \noteq f \, \tuple{t}_m) \{ x \rightarrow \lambda \tuple{y}_n. \, f \, \tuple{(z_j \,  \tuple{y}_n)}_m   \} $}  
        \singleLine
    \end{scprooftree}

     \begin{scprooftree}{1}
        \AxiomC{$C' \vee x \, \tuple{s}_n \noteq f \, \tuple{t}_m $}
        \RightLabel{\inf{Project}}
        \UnaryInfC{$(C' \vee x \, \tuple{s}_n \noteq f \, \tuple{t}_m) \{ x \rightarrow \lambda \tuple{y}_n. \, y_i \, \tuple{(z_j \,  \tuple{y}_n)}_p   \} $}  
        \singleLine
    \end{scprooftree}

\noindent
Where $j$ ranges from 1 to $m$ in \inf{Imitate} and 1 to $p$ in \inf{Project}, and each $z_j$ is a fresh variable. The literals $x \, \tuple{s}_n \noteq f \, \tuple{t}_m$ are eligible in the premises and
$p$ is the arity of $y_i$, the projected variable.
The idea behind introducing these rules
is to facilitate the instantiation of head variables
with suitable lambda terms when this is not being done 
as part of unification.
Our intuition is that by intertwining the unification
and calculus rules in the spirit of the EP calculus \cite{steen2021extensional}, the need for explosive rules 
(such as \inf{FluidSup}\cite{bentkamp2023superposition}) that
simulate superposition underneath variables is removed.
The examples we present below support this intuition.
Besides the core inference rules, the calculus has a set of rules
to handle reasoning about Boolean terms.
These are similar to rules discussed in the literature \cite{vukmirovic2020boolean,steen2018extensional}.
Extensionality is supported either via an axiom or by 
using unification with abstraction as described by Bhayat \cite{bhayat2020thesis}. 
Similarly, Hilbert choice can be supported via a lightweight 
inference in the manner of Leo-III \cite{steen2018extensional}
or via the addition of the Skolemized choice axiom.
The calculus also contains various well-known
simplification rules such as \inf{Demodulation} and \inf{Subsumption}.

\textbf{Soundness and Completeness.} The soundness of the calculus described above 
is relatively straightforward to show. 
On the other hand, the completeness of the calculus with respect
to Henkin semantics is an open question.
We hypothesise that given the right ordering,
and with tuning of inference side conditions, 
the depth$_0$ variant of the calculus (with the \inf{Imitate} and \inf{Project} rules)
is refutationally complete.
A proof is unlikely to be straightforward due 
to the fact that we do not select flex-flex literals.

\begin{example}
Consider the following unsatisfiable clause set. Assume a depth of 1.
Selected literals are underlined.

\[ C =  \underline{x \, a \, b \noteq f \, b \, a} \vee x \, c \, d \noteq f \, b \, a \]

\noindent
An \inf{EqRes} binds $x$ to $\lambda y, z. f (x_1 \, a \, b) (x_2 \, a \, b)$ and
results in 
$C_1 = \underline{f \, (x_1 \, a \, b) (x_2 \, a \, b)}$ $\underline{\noteq f \, b \, a} \vee f \, (x_1 \, c \, d) (x_2 \, c \, d) \noteq f \, b \, a$. 
An \inf{EqRes} on $C_1$ binds $x_1$ to $\lambda y, z. b$ and results in 
$C_2 = \underline{x_2 \, a \, b \noteq a} \vee f \, b \, (x_2 \, c \, d) \noteq f \, b \, a$. 
A final \inf{EqRes} on $C_2$  binds $x_2$ to $\lambda y, z. a$ and results in 
$\underline{f \, b \, a \noteq f \, b \, a}$ 
from which it is trivial to obtain the empty clause $\bot$.
\end{example}

\begin{example}[Example 1 of Bentkamp et al. \cite{bentkamp2019lhol}]
Consider the following unsatisfiable clause set. Assume the depth$_0$ version of the calculus.

\[ C_1 =  f \, a \eq c \qquad C _2  = h \, (y \, b) \, (y \, a) \noteq h \, (g \, (f \, b)) \, (g \, c) \]

\noindent
An \inf{EqRes} inference on $C_2$ results in $C_3 = y \, b \noteq g \, (f \, b) \vee y \, a \noteq g \, c$. 
An \inf{Imitate} inference on the first literal of $C_3$ followed by the application of the substitution and some $\beta$-reduction results in $C_4 = g \, (z \, b) \noteq g \, (f \, b) \vee g \, (z \, a) \noteq g \, c$. 
A further double application of \inf{EqRes} gives us $C_5 = z \, b \noteq f \, b \vee z \, a \noteq c$.
We again carry out \inf{Imitate} on the first literal followed by an \inf{EqRes} to leave us with $C_6 = x \, b \noteq b \vee f \, (x \, a) \noteq c$.
We can now carry out a \inf{Sup} inference between $C_1$ and $C_6$ resulting in $C_7 =  x \, b \noteq b \vee c \noteq c \vee x \, a \noteq a$ from which it is simple to derive $\bot$ via an application of \inf{Imitate} on either the first or the third literal. 
Note, that the empty clause was derived without the need for an inference that simulates superposition underneath variables, unlike in \cite{bentkamp2019lhol}.
\end{example}

\section{Implementation}

The calculus described above, along with a dedicated strategy schedule, has been implemented in the Vampire theorem prover.\footnote{See \url{http://bit.ly/3vBQLi4} 
for the release, \url{https://bit.ly/3Hl3lES} 
for the code.} 
Vampire natively supports rank-1 polymorphic first-order logic.
Therefore, we translate higher-order terms into polymorphic first-order
terms using the well known applicative encoding.
Note, that we use the symbol $\mapsto$, in a first-order type, to separate 
the argument types from the return type.
It should not be confused with the binary,
higher-order function type constructor $\rightarrow$ 
that we assume to be in the type signature.
Application is represented by a polymorphic symbol 
$app : \typequant \alpha_1, \alpha_2. (\alpha_1 \rightarrow \alpha_2 \times \alpha_1) \mapsto \alpha_2$. 
Lambda terms are stored internally using De Bruijn indices.
A lambda is represented by a polymorphic symbol 
$lam : \typequant \alpha_1, \alpha_2.\, \alpha_2 \mapsto (\alpha_1 \rightarrow \alpha_2)$.
De Bruijn indices are represented by a family of polymorphic symbols $d_i : \typequant \alpha. \, \alpha$ for $ i \in \mathbb{N}$.
Thus, the term $\lambda x : \tau. \, x$ is represented internally as $lam(\tau, \tau, d_0(\tau))$.
The term $\lambda x. \, f (\lambda z. x)$ is represented internally (now ignoring type arguments) as
$lam(app(f, lam(d_1)))$.

Some of the most important options available are:
\texttt{hol\_unif\_depth} to control the depth unification proceeds to,
\texttt{funx\_ext} to control how function extensionality is handled,
\texttt{cnf\_on\_the\_fly} to control how eager or lazy the clausification algorithm is, and
\texttt{applicative\_unif} which replaces higher-order unification with (applicative) first-order 
unification. 
This is surprisingly helpful in some cases.
Besides for the options listed above, there are many other higher-order specific
options as well as options that impact both higher-order
and first-order reasoning. 
These options can be viewed by building Vampire and running with \texttt{--help}.

\section{Strategies and the Schedule}

We generally followed the Spider \cite{Spider} methodology for strategy discovery and schedule creation.
This starts with randomly sampling strategies to solve as-of-yet unsolved problems 
(or improve the best known time for problems already known to be solvable). 
Each newly discovered strategy is optimized with local search to work even better on
the single problem which it just solved. This is done by trying out alternative values for each option,
possibly in several rounds. A variant of the strategy that improves the solution time or at least uses a default value of an option is preferred. 
The final strategy is then evaluated on the totality of all considered problems and the process repeats.


In our case, we sought strategies to cover the \num{3914} TH0 problems of the TPTP library \cite{Sut17} version 8.1.2.
The strategy space consisted of 87 options inherited from first-order Vampire and 26 dedicated higher-order options.
%
%
To sample a random strategy, we considered each option separately and picked its value based on a guess of how useful each is. 
(E.g., for \texttt{applicative\_unif} we used the relative frequencies of \texttt{on}: 3, \texttt{off}: 10.)
During the strategy discovery process we adapted the maximum running time per problem,
both for the random probes several times and for the final strategy evaluation: from the order of \SI{1}{\second} up to \SI{100}{\second}.
%
%
%
In total, we collected \num{1158} strategies over the course of approximately two weeks of continuous 60 core CPU computation.
The strategies cover \num{2804} unsatisfiable problems, including 50 problems of TPTP rating 1.0 
(which means these problems were not officially solved by an ATP before).

Once a sufficiently rich set of strategies gets discovered and evaluated, 
schedule building can be posed as a constraint programming task
in which one seeks to allot time slices to individual strategies 
to cover as many problems as possible while not exceeding a given overall time bound $T$ \cite{DBLP:conf/mkm/HoldenK21,DBLP:conf/paar/Schurr22}.
We had a good experience with a weighted set cover formulation and applying a greedy algorithm \cite{DBLP:journals/mor/Chvatal79}:
starting from an empty schedule, at any point we decide to extend it by scheduling a strategy $S$ 
for additional $t$ units of time if this step is currently the best among all possible strategy extensions in terms of 
``the number of problems that will additionally get covered \emph{divided by} $t$''. This greedy approach does not guarantee
an optimal result, but runs in polynomial time and gives a meaningful answer uniformly for any 
overall time bound $T$. (See \cite{snakeAccepted} for more details).

Our final higher-order schedule tries to cover, in this greedy sense, as many problems 
as possible at several increasing time bounds: 
starting from \SI{1}{\second}, \SI{5}{\second}, and \SI{10}{\second} bounds relevant for the impatient users,
all the way up to the CASC limit of 16 minutes (2 minutes on 8 cores) and further beyond.
In the end, it makes use of \num{278} out of the \num{1158} available strategies
and manages to cover all the known-to-be-solvable problems in a bit less than 1 hour of single core computation.
We stress that our final schedule is a single monolithic sequence
and does not branch based on any problems' characteristics or features.\footnote{One additional interesting aspect of our schedule building approach
(left out due to space restrictions, but see Appendix~\ref{sect:prob_sched_app} for more details
) is that we employ input shuffling and prover randomization \cite{DBLP:conf/cade/Suda22} and
thus treat our strategies as Las Vegas algorithms, whose running time or even success/failure may depend on chance.}


\paragraph{Most important options:}

\begin{table}
\centering
\caption{The most important options in terms of contribution to problem coverage}
\label{tab:important_options}
\begin{tabular}{lcc}
an option & default & \# problems not solvable without non-default \\
\hline
\texttt{cnf\_on\_the\_fly} & \texttt{eager} & 102 \\
\texttt{applicative\_unif} & \texttt{off} & 56 \\
\texttt{equality\_to\_equiv} & \texttt{off} & 24 \\
\texttt{hol\_unif\_depth} & \texttt{2} & 20 \\
\texttt{func\_ext} & \texttt{abstraction} & 12 \\
\end{tabular}
\end{table}

In Table~\ref{tab:important_options}, we list the first five options sorted in descending order of ``how many problems 
we would not be able to cover 
if the given option could not be varied in strategies.'' (In other words, as if the listed default value was ``wired-in'' to the prover code.)

Based on existing research \cite{vukmirovic2021making},
it is unsurprising to see that varying clausification has a large impact.
Likewise, for varying the unification depth.
What is perhaps more surprising is that replacing higher-order
unification with applicative first-order unification 
can be beneficial.
\texttt{equality\_to\_equiv} turns equality between Boolean terms
into equivalence before the original clausification pass is
carried out.
The effectiveness of this option is also somewhat surprising.


\paragraph{Performance statistics:}

\begin{table}
\centering
\caption{Number of problems solved by a single good higher-order strategy and our schedule at various time limit cutoffs.
Run on the \num{3914} TH0 TPTP problems}
\label{tab:strat_vs_sched}
%
%
%
%
\setlength{\tabcolsep}{8pt}
\begin{tabular}{l|cccccc}
 & \SI{1}{\second} & \SI{10}{\second} & \SI{30}{\second} & \SI{60}{\second} & \SI{120}{\second} & \SI{960}{\second} \\
\hline
single strategy & 1811 & 1949 & 2041 & 2094 & $-$ & $-$ \\
our schedule    & 2067 & 2436 & 2584 & 2642 & 2691 & 2775 \\
\end{tabular}
\end{table}

It is long known \cite{DBLP:conf/cade/Tammet98,DBLP:conf/flairs/WolfL98} that a strategy schedule can improve 
over the performance of a single good strategy by large margin.
Table~\ref{tab:strat_vs_sched} confirms this phenomenon for our case. 
For this comparison we selected one of the best performing (at the \SI{60}{\second} time limit mark) single strategies 
that we had previously evaluated.
From the higher-order perspective, the strategy is interesting for setting \texttt{hol\_unif\_depth} to 4 and supporting choice reasoning via an inference rule (\texttt{choice\_reasoning on}).
\footnote{Otherwise, it uses Vampire's default setting, except for relying on an incomplete literal selection function \cite{DBLP:conf/cade/HoderR0V16}
and using a relative high naming threshold \cite{DBLP:conf/gcai/Reger0V16}, i.e., being reluctant to introduce new names for subformulas during clausification.}

Although our schedule has been developed on (and for) the TH0 TPTP problems, 
it helps the new higher-order Vampire solve more problems of other origin too.
%
%
%
Of the Sledgehammer problems exported by Desharnais et al.~in their last prover comparison \cite{DBLP:conf/itp/DesharnaisVBW22},
namely the 5000 problems denoted in their work TH0$^-$, Vampire can now solve \num{2425}
compared to \num{2179} obtained by Desharnais et al.~with the previous Vampire version (both under \SI{30}{\second} per problem).\footnote{
Our experiments were run on Intel\textregistered Xeon\textregistered Gold 6140 CPU @ \SI{2.3}{\giga\hertz}, 
Desharnais et al.~\cite{DBLP:conf/itp/DesharnaisVBW22} used StarExec \cite{DBLP:conf/cade/StumpST14} with Intel\textregistered Xeon\textregistered CPU E5-2609 @ \SI{2.4}{\giga\hertz} nodes.}

We remark that we also developed a different schedule 
specifically adapted to Sledgehammer problems
(in various TPTP dialects, i.e., not just TH0), which is now available to the Isabelle \cite{nipkow2002isabelle} users since the September 2023 release.








\section{Related Work}

The idea to intertwine superposition and unification appears in earlier work, particularly in the EP calculus implemented in Leo-III \cite{steen2021extensional}.
The main differences between our calculus and EP are:
\begin{enumerate}
\item We do not move first-order unification to the calculus level. Hence, there are no equivalents to the \inf{Triv}, \inf{Bind} and \inf{Decomp} rules of EP.
\item Our \inf{Project}  and \inf{Imitate} rules
are instances of EP's \inf{FlexRigid} rule.
We do not include an equivalent to 
EP's \inf{FlexFlex} rule since we never select flex-flex literals.
Instead, we leave such literals until one
of the head variables becomes instantiated,
or the clause only contains flex-flex literals
at which point \inf{FlexFlexSimp} can be applied.
\item Our core inference rules are
parameterised by a selection function and an ordering.
\item Whilst EP always applies unification lazily, our calculus can control how lazily 
unification is carried out by varying the depth bound.\footnote{Our understanding based on \cite{steen2018extensional} is that the implementation of EP in Leo-III does make use of orderings as well as eager unification. It is not clear to us from the exposition how these are applied. }
\end{enumerate}
We also incorporate more recent work on higher-order superposition,
mainly from the Matryoshka project \cite{vukmirovic2021making,bentkamp2023superposition}.
Of course, the use of constraints in automated reasoning
extends far beyond the realm of higher-order logic.
They have been researched in the context of theory reasoning \cite{reger2018unification,korovin2023alasca} and
basic superposition \cite{bachmair1992basic}. 

\section{Conclusion}

In this paper, we have presented a new higher-order superposition 
calculus and discussed its implementation in Vampire. 
We have also described the new higher-order schedule created.
The combination of calculus, implementation and schedule 
have already proven effective.
However, we believe that there is great room for
further exploration and improvement.
On the theoretical side, we wish to prove refutational completeness of the calculus (or a variant thereof).
On the practical side, we wish to refine the implementation, most notably by adding
additional simplification rules.

\begin{credits}
\subsubsection{\ackname}
The second author was supported by
project CORESENSE no.~101070254 under the Horizon Europe programme
and
project RICAIP no.~857306 under the EU-H2020 programme.
\end{credits}

\bibliographystyle{splncs04}
\bibliography{main}

\newpage

\appendix
\section{Note on Probabilistic Schedule Building}
\label{sect:prob_sched_app}

CASC organizers randomly shuffle the input problems to make sure that ``no system receives an advantage or disadvantage 
due to the specific presentation'' \cite{CASC2023}. 
At the same time, it is well known that with a saturation-based prover even such small changes 
may have dramatic effect on strategy's running time \cite{DBLP:conf/cade/Suda22}. 
To create a schedule resilient to input shuffling,
we actually sampled our strategies with Vampire's internal shuffling enabled,
ready to evaluate a strategy on a problem more than once (under different random seeds) 
to establish an estimate of its runtime distribution. 

We then adapted the greedy algorithm to seek to cover problems ``in expectation''. By this we mean that 
a strategy can score a fractional point for solving a problem (if it solves it, e.g., in \SI{50}{\percent} of its runs),
where appropriately smaller fractions are awarded for problems already partially covered.

Our final schedule is then also executed under fresh random seeds and its performance, therefore,
slightly varies depending on chance.

The estimate of a runtime distribution of a strategy on a given problem 
is explained on an example in Table~\ref{tab:estimate}. Note that we work
with two possible failure modes: a deliberate giving up, which may happen with incomplete strategies,
and termination through a timeout. The difference in interpretation
is that a timeout could later have turned into a success if we had waited longer.

\begin{table}
\caption{Four independent example runs of a strategy on a given problem.
The probability estimate for solving the given problem at intervals $I_1,\ldots,I_5$ 
changes between time moments $t_1,\ldots,t_4$ where one of the runs changes status from running (r) 
to either success (i.e, solved), timeout (solution interrupted), or gave up (premature failure). }
\label{tab:estimate}
\centering
\begin{tabular}{c|cccc|c}
time points and intervals & run$_1$ & run$_2$ & run$_3$ & run$_4$ & probability \\
\hline
$I_1$ &    r    & r & r & r & 0/4 \\
$t_1$ & success &     r    &    r     &     r    &     \\
$I_2$ &         & r & r & r & 1/4 \\
$t_2$ &         & gave up & r & r &     \\
$I_3$ &         &         &    r     &    r     & 1/4 \\
$t_3$ &         &         & timeout & r &     \\
$I_4$ &         &         &         &    r    & 1/3 \\
$t_4$ &         &         &         & success &     \\
$I_5$ &         &         &         &         & 2/3 \\
\end{tabular}
\end{table}

We initially evaluate each strategy only once, as described in the main text. We then run 
the greedy schedule construction algorithm multiple times and iteratively reevaluate strategies on problems
which the greedy algorithm reports are getting covered by them, thus getting gradually better probability estimates
at points where they matter. 

Covering problems in expectation uses the assumption of strategy independence. For example, 
knowing that the current schedule solves problem $P$ with a probability of 0.6
and a new strategy $S$ added for $t$ units of time solves $P$ with the probability \num{0.8}, adding $S$ to the schedule will 
improve the coverage of $P$ from 0.6 to \num{0.92}, that is, by $r = (1-0.6) \cdot 0.8$.
When computing the weight for the greedy covering,
the strategy will accrue these $r$ points for contributing to solving problem $P$.

\begin{figure}
\caption{A ``cactus'' plot comparing the performance of the probabilistic schedule with a deterministic one. 
Run on (upper) the \num{3914} TH0 problems from the TPTP library version 8.1.2 (training data) 
and on (lower) the 5000 TH0$^-$ problems from Desharnais et al.~\cite{DBLP:conf/itp/DesharnaisVBW22} (unseen during schedule construction). The time limit was \SI{960}{\second} per problem}
\label{fig:plot}
\centering
\includegraphics[scale=0.8]{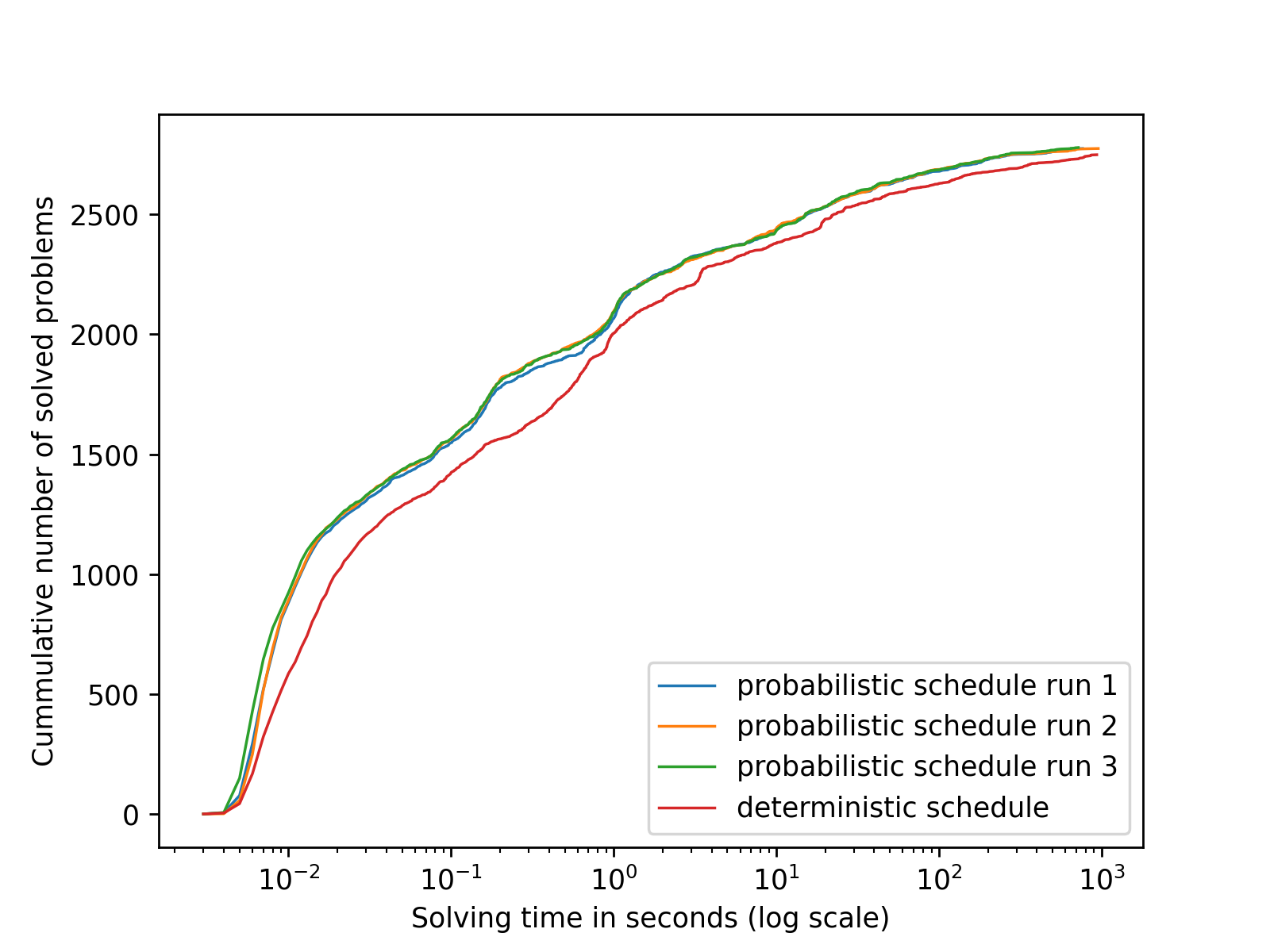}
\includegraphics[scale=0.8]{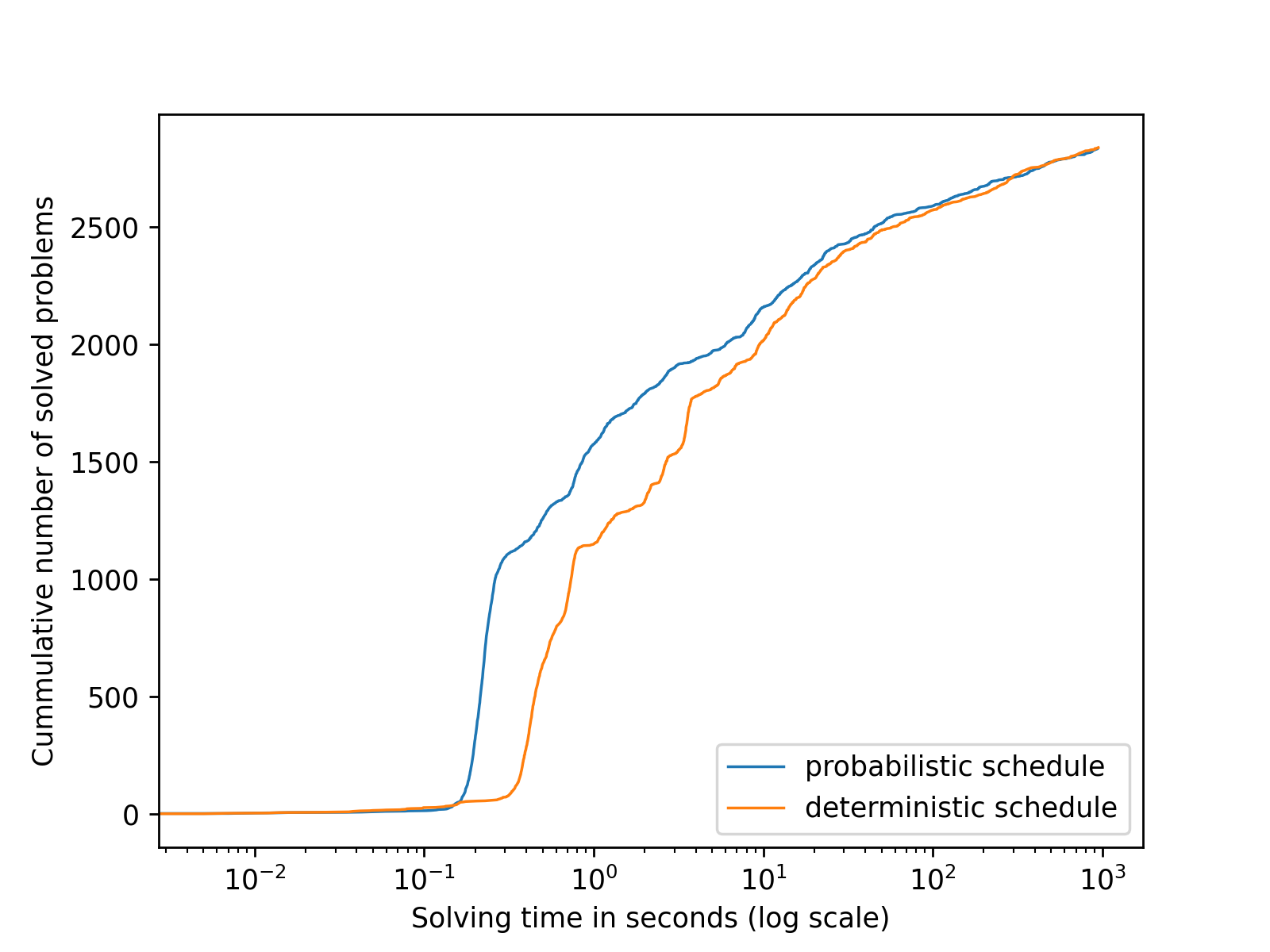}
\end{figure}

The performance impact of the probabilistic approach can be observed in Figure~\ref{fig:plot}.
We can see that there is a substantial difference in the lower time limit bracket 
(around 90 problems at \SI{1}{\second} on TPTP and 426 problems on the Sledgehammer problems),
which becomes less pronounced with higher time limits (50 problems at \SI{30}{\second} on TPTP and
31 on Sledgehammer there). The gap completely closes at around \SI{300}{\second} per problem
on the Sledgehammer problems, from which point the deterministic schedule even mildly dominates.
We currently do not have a good explanation for this last observation.



We conclude that investing into developing a probabilistic schedule on higher-order problems does pay off,
especially at low time limits desired in applications with an impatient user,
and carries over to in-training-unseen problems.

\end{document}